\documentclass{elsart}

\usepackage{graphicx}
\newcommand{\be}{\begin{eqnarray}}
\newcommand{\ee}{\end{eqnarray}}
\usepackage{amssymb}

\begin{document}

\begin{frontmatter}



\title{The conical flow from quenched jets in sQGP}


\author{E.Shuryak}

\address{ Department of Physics and Astronomy, University at Stony Brook,\\
Stony Brook NY 11794 USA}

\begin{abstract}
Starting with a reminder of what is strongly coupled Quark-Gluon
Plasma
(sQGP), we proceed to recent advances in jet quenching and heavy quark
diffusion, with a brief summary of various results based on AdS/CFT
correspondence. The conical flow is a hydrodynamical phenomenon
created by energy and entropy deposited by high energy jets
propagating in matter, similar in nature to well known sonic boom
from the supersonic planes. After a brief review, we discuss
excitations  of two hydro modes -- sound and ``diffuson'' --
which can be excited in this way. We also study expanding matter case,
with a variable sped of
sound,
and  use adiabatic
invariants to show that the parameter $v/T$ ($v$ velocity in the wave, $T$
temperature) is increasing, up to a factor 3, during expansion.
At the end we discuss recent results of the Princeton group
which derived conical flow from AdS/CFT.
\end{abstract}

\end{frontmatter}

\section{Why do we think that QGP is strongly coupled?}

 A realization~\cite{Shu_liquid,SZ_rethinking,SZ_CFT} that
QGP at RHIC  is  a strongly coupled liquid
 has lead to a  paradigm
 shift in the field. It was extensively debated  at
the ``discovery''  BNL workshop in 2004~\cite{discovery_workshop} (at which the
abbreviation sQGP was established) and multiple other meetings since.
In the intervening three years we all
had to learn a lot, in part 
 from the branches of physics  which
happened to have  experience with strongly coupled systems.
Those range from
quantum gases to classical plasmas to string theory.
In short, there seem to be not one
but  actually two difficult
issues we are facing. One is to understand why QGP at $T\sim 2T_c$
is strongly coupled, and what exactly it means. The second large
problem is to understand what  happens
 at the deconfinement, at $|T-T_c| \ll T_c$,
which may be a key to the famous confinement problem. 

As usual, progress proceeds from
 catching/formulating the main concepts and qualitative pictures, to
mastering technical tools, to final quantitative
predictions: and now we are somewhere in the middle
of this process. The work is going on at many fronts.
At classical level, 
 first studies of the  transport properties
of strongly coupled non-Abelian plasmas have been made.
 Quantum-mechanical studies
of the bound states above $T_c$ have revealed a lot of unusual
states, including ``polymeric chains''.  At the quantum field/string 
theory front, a surprisingly detailed uses of
 AdS/CFT correspondence has been made (see below).

The list of arguments  why we think  QGP
is strongly coupled at $T=(1-2)T_c$
is long and  growing. Let me start with its
short version, as I see them today.\\
\noindent
1.Collective phenomena observed at RHIC lead
 hydro practitioners to a conclusion that
 QGP as a ``near perfect liquid'', with 
small {\em viscosity-to-entropy ratio}
$\eta/s=.1-.2<<1$ \cite{Teaney:2003kp}.
Not only soft partons rescatter a lot, but even
high energy jets, including the charmed ones, are strongly quenched.
Even in this case one needs hydrodynamics to tell what is happening
(conical flow).
\\
2. Combining lattice data on quasiparticle masses and
inter-particle
potentials, one  finds a lot of quasiparticle bound states
\cite{SZ_bound}. 
This approach explains why $\eta_c,J/\psi$ remains bound
till  $T<(2-3)T_c$, as was directly shown on the lattice
\cite{charmonium} and perhaps experimentally at RHIC.
The bound states and resonances
enhance transport cross sections \cite{SZ_rethinking,Rapp_vanHees}
which helps to explain liquid-like behavior. Similar phenomena are known
 for ultracold trapped atoms, which can be put in a liquid 
form via
Feshbach resonances at which
the scattering length $a\rightarrow \infty$.
\\
3.Classical interaction parameter  
$\Gamma\sim <potential\, energy>/<kinetic\, energy>$
in sQGP is not small at all. Classical
  e/m plasmas  at the comparable
coupling $\Gamma\sim 1-10$ are known to be good liquids.
Our results \cite{GSZ} show it to be also true for non-Abelian
plasma as well.
\\
4. Correspondence between the conformal
(CFT) $\mathcal N$=4 supersymmetric Yang-Mills theory
at strong coupling and string theory in Anti-de-Sitter
space (AdS) in classical SUGRA regime was conjectured by Maldacena~\cite{Maldacena:1997re}.
 The results obtained this way on the 
 $g^2N_c\rightarrow \infty$ regime of the CFT plasma 
 are all  close to what we
 know about sQGP (see below).
\\
5.
The $\mathcal N$=2 SUSY YM (``Seiberg-Witten'' theory) 
is a working example of confinement due to 
condensed monopoles\cite{SW}.
If it is also true for QCD,  at $T\rightarrow T_c$
magnetic monopoles  become light
 and weakly interacting at large distances due to $U(1)$ beta function.
Then the Dirac condition forces
 electric coupling $g$ to become large (in IR).

This list is full  of fascinating issues, but 
most of them are $not$ what the HP06
organizers  asked me to speak about, which is
a  focus on  ``hard probe''
part of the story, to which I now proceed.

\section{Jet quenching and heavy quark diffusion}
\label{}
A decade ago,  at the formation time of RHIC program, the ``hard probe''
community clearly looked from above at the ``soft probes''.
Indeed, the letter spoke about the phase diagram,
temperature and entropy, flow and viscosity and other macroscopic
quantities which was quite difficult to justify for not-too-large
systems we have.  ``Hard'' physics was much simpler conceptually, and
had a  well calibrated tool -- pQCD and a parton model PDFs. 
``Calibration'' was the key word, and many small effects 
(like gluon shadowing) got a lot of attention.

And yet, what happened at RHIC was quite contrary to these  expectations:
  hydrodynamics turned out to be the $only$ theory which
actually worked at RHIC! It 
provided quantitative description of  spectra of most secondaries\footnote{Well, only
 about 99 percent, for $p_t< 2 GeV$.}. ``Parton cascades''
 had spectacularly flopped, unable to generate any
 collective
flow, unless their pQCD cross sections are grossly enhanced.

  Then came the jet quenching  $R_{AA}$ data, with a shocking conclusion
that we can only see 
only small fraction of all jets, perhaps originating in a dilute corona.
With  recent data on charm (single electrons)
 hopes to describe it 
 by the gluon radiation via
pQCD are fading. Although these issues are
discussed at HP06 extensively, let me still mention
few main reasons for this conclusion.\\
(i)
The quenching of charm quarks turned out to be very similar
to that of the usual jets without charm. First of all, this
contradicts the Casimir scaling of any perturbative diagram, which
would demand that a (charmed) quark has a smaller charge than a gluon
(making
most of the usual jets at RHIC. \\(ii)  Radiative  jet
quenching should be reduced for heavy quarks: thus now a discussion of 
collisional loss restarted, perhaps enhanced by heavy-light
resonances \cite{Rapp_vanHees}.Zahed and myself\cite{Shuryak:2004ax}
have also calculated ``ionization'' energy loss due to binary bound states.\\
 (iii) Last but not least: secondaries from the
quenched jets are not at all emitted in the direction of a jet but
rather
at very large angle $\sim 70^o$ to it. It is hard to see how
 a gluon radiation may show such pattern: but  it is exactly what
the hydrodynamical theory of sound emission is predicting. 

Charm diffusion and drag can be described by a single
constant, $D_c$, which can be extracted from comparison to charm $R_{AA}$
and charm elliptic flow, see e.g. Moore and Teaney \cite{MT}.
The resulting phenomenological value from RHIC data
is in the range\be \label{eqn_Dc}
D_c*(2\pi T)=(1-2) \ee
while the pQCD result is 
\be  D^{pQCD}_c*(2\pi T)=1.5/\alpha_s^{2} \ee
 Assuming that perturbative domain is\footnote{Recall
that at $4/3\alpha_s=1/2$ two scalar quarks should fall at each other,
according to Klein-Gordon eqn: so this is clearly not a perturbative region.} 
somewhere at $\alpha_s< 1/3$ one concludes that empirical value
(\ref{eqn_Dc}) is  an order of
magnitude smaller than the perturbative 
one.

\section{AdS/CFT and strongly coupled plasma: brief summary}
\label{}

With  weak  coupling methods failing, one  gets interested
 in the  strong coupling limit. AdS/CFT correspondence
is one of the directions (but not the only one!)
which allows to address it,
so far not for QCD but for  its distant
cousin $\mathcal N$=4 SYM theory. 

{\bf Thermodynamics} of the CFT plasma was studied started from the
early work\cite{thermo}, its  result 
is that the free energy (pressure) of a plasma is
\be F(g,N_c,T)/ F(g=0,N_c,T)=[(3/4)+O((g^2N_c)^{-3/2})] \ee 
which  compares well with the
 lattice value\footnote{ Not too close to $T_c$, of course, but
 in the
``conformal domain'' of $T= few\,T_c$, in which
$p/T^4$ and $\epsilon/T^4$ are constant.} of about $0.8$.

{\bf Heavy-quark potentials} in vacuum  and then
at finite $T$ \cite{AdS_pot}
 were calculated by calculating the configuration
of the static string, deformed by gravity into the 5-th dimension.
Let me write the result schematically as
\be V(T,r,g)\sim  {\sqrt{g^2 N_c}\over r} exp(-\pi T r)  \ee 
The
 Debye radius, unlike in pQCD , has no
 coupling constant: but it is not far from lattice value at $\sim 2 T_c$.
Although potential depends on distance $r$ still
as in the Coulomb law, $1/r$ (at $T=0$ it is due to conformity),
 it is has a notorious square root of the coupling. 
Semenoff and Zarembo \cite{Semenoff:2002kk} noticed that summing ladder diagrams
one can explain $\sqrt{g^2 N_c}$, although
not a numerical constant. Zahed and myself~\cite{SZ_CFT}
pointed out that  both static charges are color correlated 
during a parametrically small time  $\delta t\sim r/{(g^2
  N_c)^{1/4}}$: this explains~\cite{Klebanov:2006jj} why a field of the dipole   
is $1/r^7$ at large distance\cite{Callan:1999ki}, not  $1/r^6$. 
Debye screening range can also be explained by resummation of thermal
polarizations~\cite{SZ_CFT}.

 Zahed and myself~\cite{SZ_spin} had also discussed
the velocity-dependent
forces , as well as spin-spin and spin-orbit ones, at strong coupling.
Using
ladder resummation for non-parallel
Wilson lines with spin they  concluded that all of them
join into one common square root
\be V(T,r,g)\sim \sqrt{(g^2 N_c)[1-\vec v_1 *\vec v_2+(spin-spin)+(spin-orbit)]}/ r   \ee 
Here $\vec v_1,\vec v_2$ are velocities of the quarks: 
and the corresponding term is a strong coupling version of Ampere's
interaction between two currents\footnote{Note that in a 
quarkonium  their scalar product is negative,
increasing attraction.}. No results on that are known from a gravity
side, to my knowledge.

{\bf Bound states:} Zahed and myself~\cite{SZ_CFT} looked for
 heavy quarks   bound states, using a  Coulombic potential with 
Maldacena's $\sqrt{g^2 N_c}$ and
 Klein-Gordon/Dirac eqns. There is no problem with states
 at large orbital momentum $J>>\sqrt{g^2N_c}$, otherwise one has 
the famous  ``falling on a center'' solutions\footnote{Note that all
 relativistic corrections mentioned above  cannot prevent it
from happening.}: we argued that
a significant density of bound states develops, at all energies, from zero to
$2M_{HQ}$.  
And yet, a study of the gravity side~\cite{Kruczenski:2003be} 
found that there is no falling. In more detail, the
Coulombic states at large $J$ are supplemented by two more
families: Regge ones with the mass $\sim M_{HQ}/(g^2N_c)^{1/4}$ and
 the lowest $s$-wave states
(one may call $\eta_c,J/\psi$)  with even smaller
masses  $\sim M_{HQ}/\sqrt{g^2  N_c}$.
 
The issue of ``falling'' was further discussed
 by Klebanov, Maldacena and Thorn~\cite{Klebanov:2006jj} for
 a pair of static quarks: they calculated the
 spectral density of states
  via a semiclassical quantization of  string vibrations.
They argued  that their corresponding density of states
should appear at exactly the same
critical
coupling as the famous ``falling'' in the Klein-Gordon eqn..

AdS/CFT also has  multi-body states  similar
to   ``polymeric
chains'' $\bar q.g.g... q$ discussed above. For
the endpoints being static quarks and the intermediate
gluons  conveniently replaced by adjoint scalars,
 Hong, Yoon and Strassler~\cite{Hong:2004gz} have studied
such states and even their formfactors.

{\bf Transport properties} of the CFT plasma
have been pioneered
  by Polikastro, Son and Starinets\cite{PSS} who
 have calculated  viscosity
(at infinite coupling). Their famous result
 $ \eta/s=>1/4\pi$
 is again in the ballpark of
 the empirical RHIC value. 
Thus
$gravitons$ in the bulk  are dual
to $sound$ on the brane. Dual to
 (viscous) sound absorption is  interception
of gravitons by the black hole.

{\bf Heavy quark diffusion constant}
has been calculated by Casalderrey and
Teaney~\cite{Casalderrey-Solana:2006rq}: their result 
\be D_{HQ}={2 \over \pi T \sqrt{g^2 N_c}}  \ee
 is even parametrically smaller than 
the expression for  momentum diffusion
$D_p=\eta/(\epsilon+p)\sim 1/4\pi T$.
If one plug in numbers, one can get $D_c*(2\pi T)\sim 1$,
in the RHIC ballpark again.

\begin{figure}
\includegraphics[width=5cm]{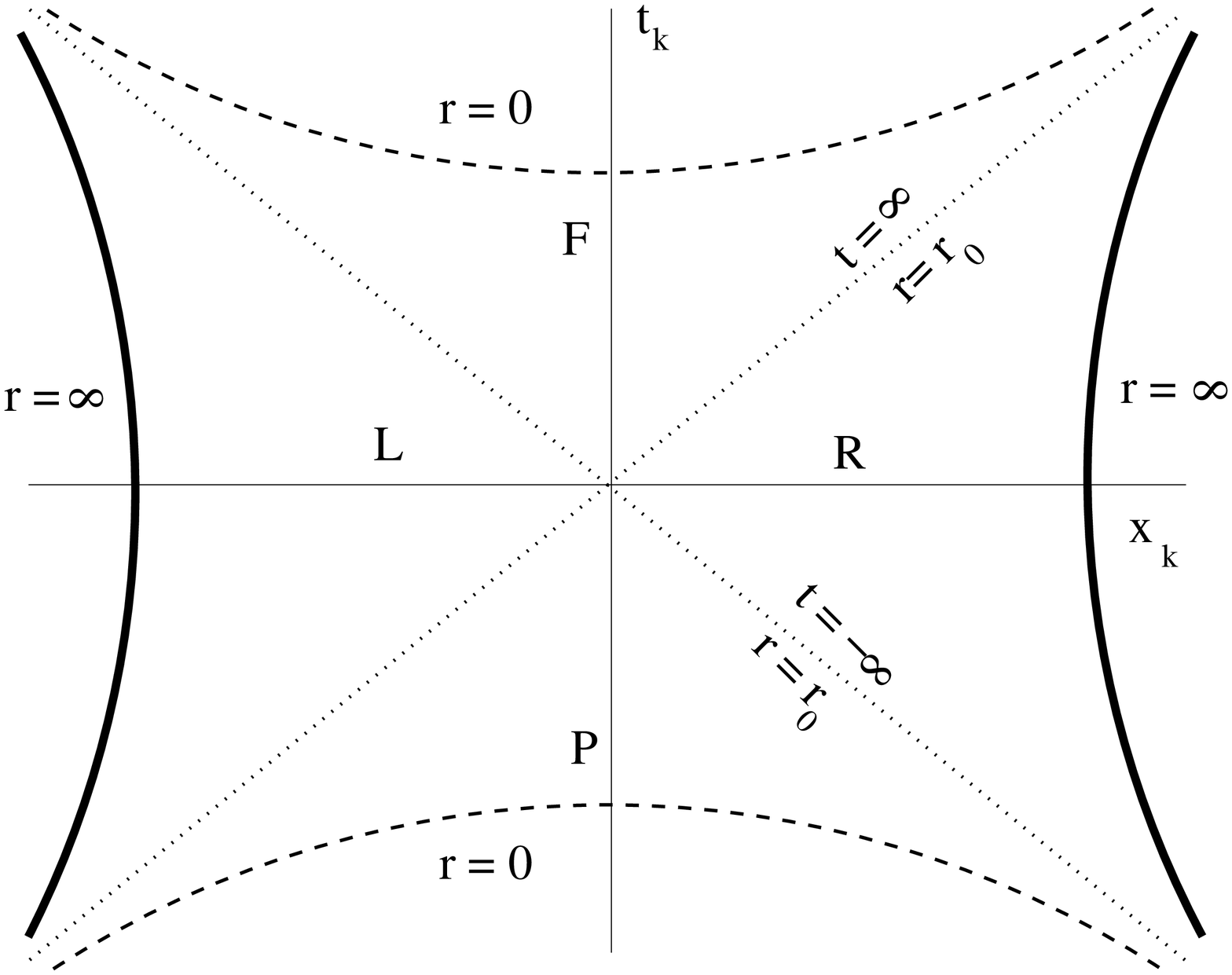}
\includegraphics[width=6cm]{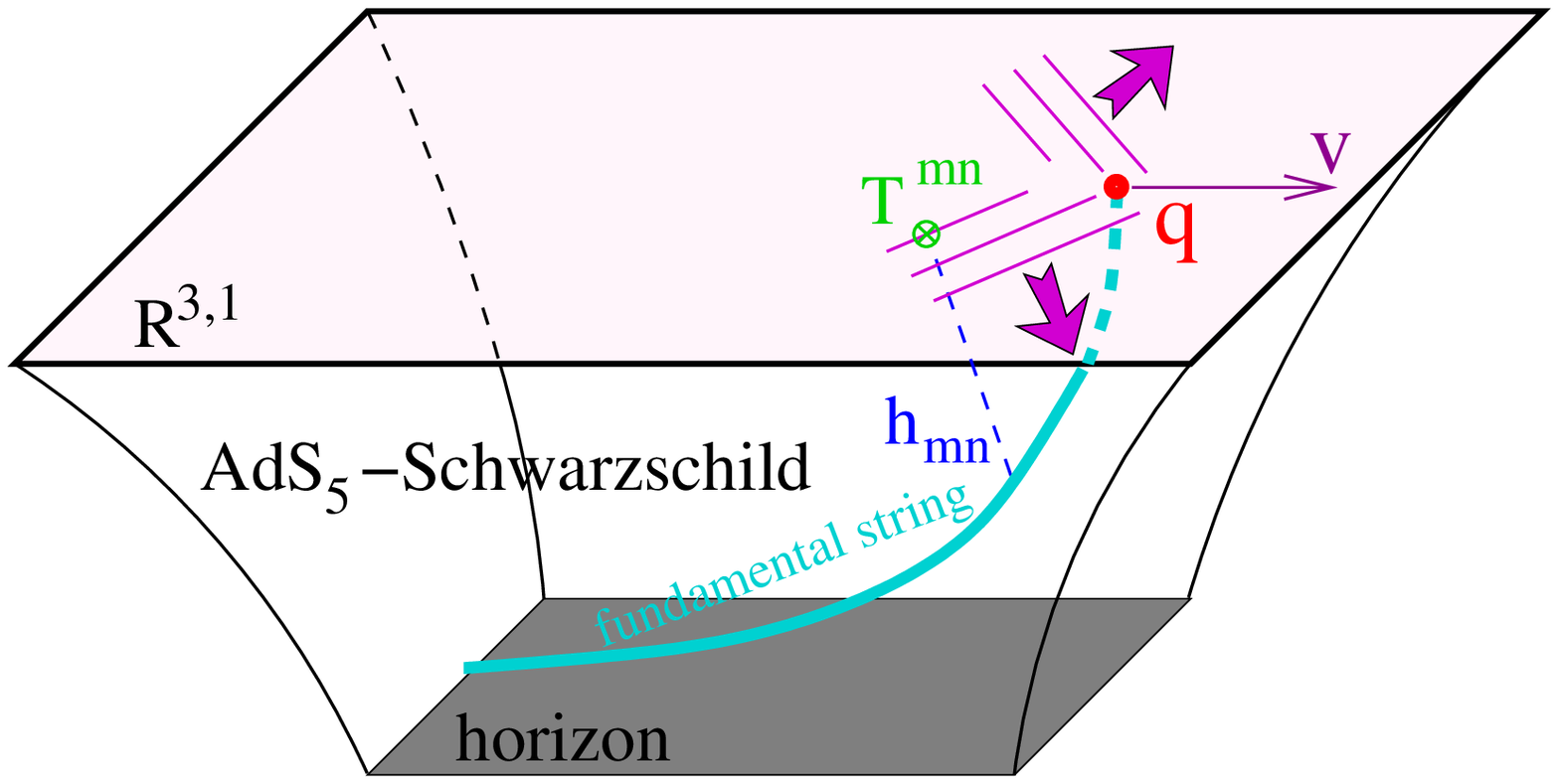}
 \caption{\label{fig_wake}
(a) (from \protect\cite{Casalderrey-Solana:2006rq}): In Kruskal coordinates one can study two Universes
at the same time, shown right and left, and the evaluated
Wilson line contains static quarks on their boundaries.
(b) (from \protect\cite{Friess:2006aw})
The dragged quark trails a string into the five-dimensional AdS
 bulk, representing color fields sourced by the quark's fundamental
 charge and interacting with the thermal medium. The back gravity
 reaction describes how matter flows on the brane.
 }
 \end{figure}

Note that this important work is methodically quite different from others in that Kruskal
coordinates are used, which allows to consider the inside
of the black hole and $two$ Universes (with opposite time directions)
simultaneously, see Fig.\ref{fig_wake}a. This is indeed necessary
\footnote{One such problem is evaluation of the so called $\hat q$
parameter: two lines of the loop  should 
also belong to $two$ different Universes, not one as assumed in
\cite{Liu:2006ug}. It remains unknown  whether
similar calculation in Kruskal geometry
would produce the same result or not. }
 in any problems when
a $probability$ is evaluated, because that contains
both an amplitude and
a conjugated amplitude at the same time. 
  
{\bf Heavy quark quenching}
\cite{Sin:2004yx,Herzog:2006gh,Gubser:2006bz,Buchel:2006bv,Sin:2006yz}
 for quarks heavy enough $M>M_{eff}\sim
\sqrt{g^2 N_c} T$  is
obtained in a stationary setting, in which
a quark is dragged with constant by ``an invisible hand'' via some rope
through QGP, resulting in constant production of a string length per
time, see Fig.\ref{fig_wake}(b). The 
resulted  drag force 
\be {dP\over dt}= -{\pi T^2\sqrt{g^2 N_c} v \over  2\sqrt{1-v^2}} \ee
 remarkably satisfies the Einstein relation which 
relates it the heavy quark diffusion constant (given above),
 in spite of quite different gravity settings.

\section{Brief history of the conical flow}
\label{conical}
  In 1980's  Greiner and collaborators
discussed possible shock wave formation and Mach cone emission
in light-on-heavy collisions at BEVALAC. It did not work, because
nuclear matter is $not$ a good liquid. 
 
  Since we now known sQGP $is$  a very good liquid, 
 Casalderrey, Teaney and myself suggested\footnote{To be fair,
this idea only came to our mind $after$ we have seen the
first STAR and PHENIX data on angular correlations.} \cite{CST} that
 {\em conical} flow must  be induced by
jets. We argued that as
 jet dumped  energy into the medium locally, 
it should partially be transformed into  coherent radiation of
sound waves.  Unlike gluons\footnote{This argument 
is one of several good
reasons to discard 
 Cerenkov gluon radiation as a source of conical distributions.
},
they are  much less absorbed
 and can propagate  till the end (freezeout) and  be detected.
(In fact we will argue below that their amplitude 
should even grow with time.)

\begin{figure}
 \includegraphics[width=7cm]{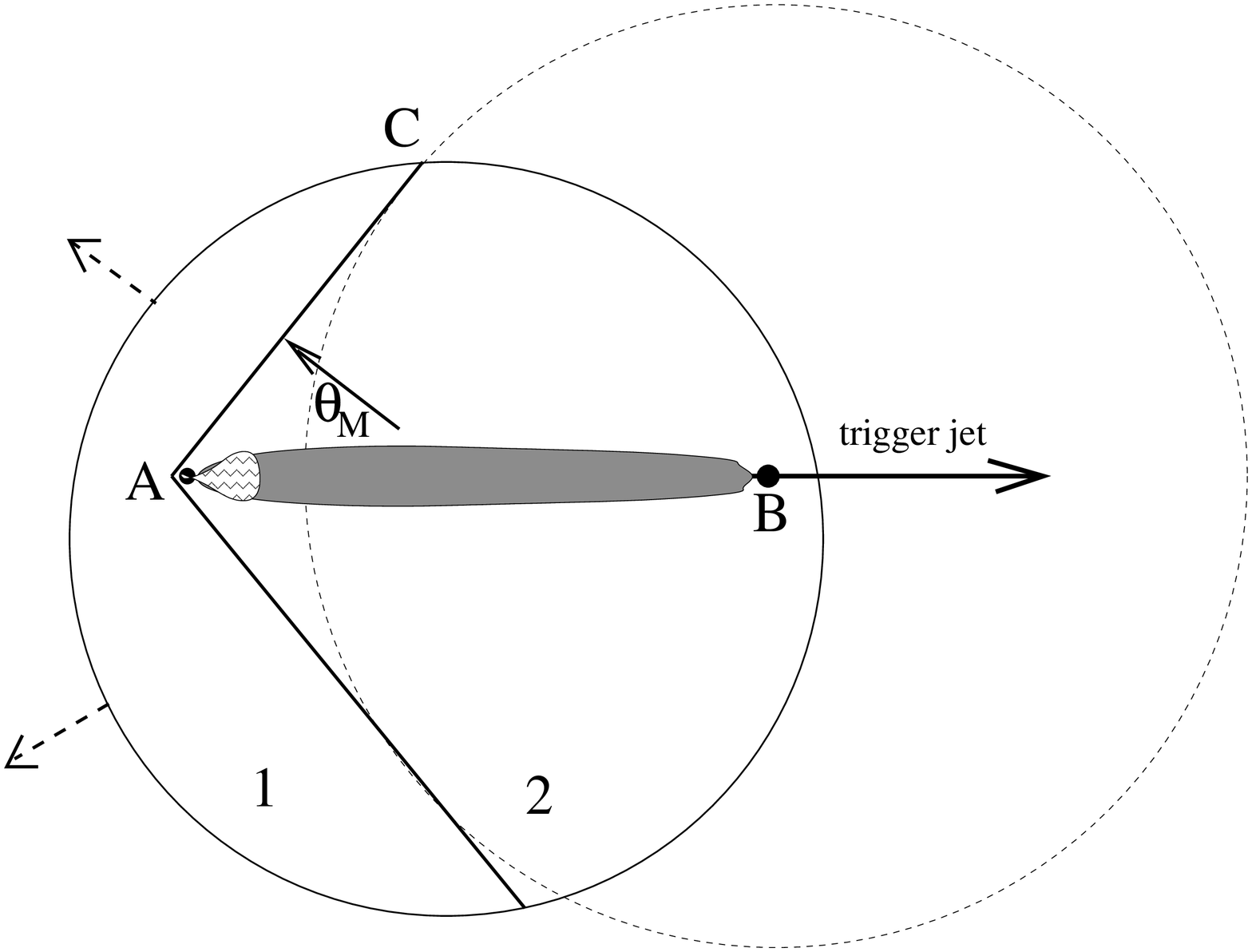}
 \includegraphics[width=5cm]{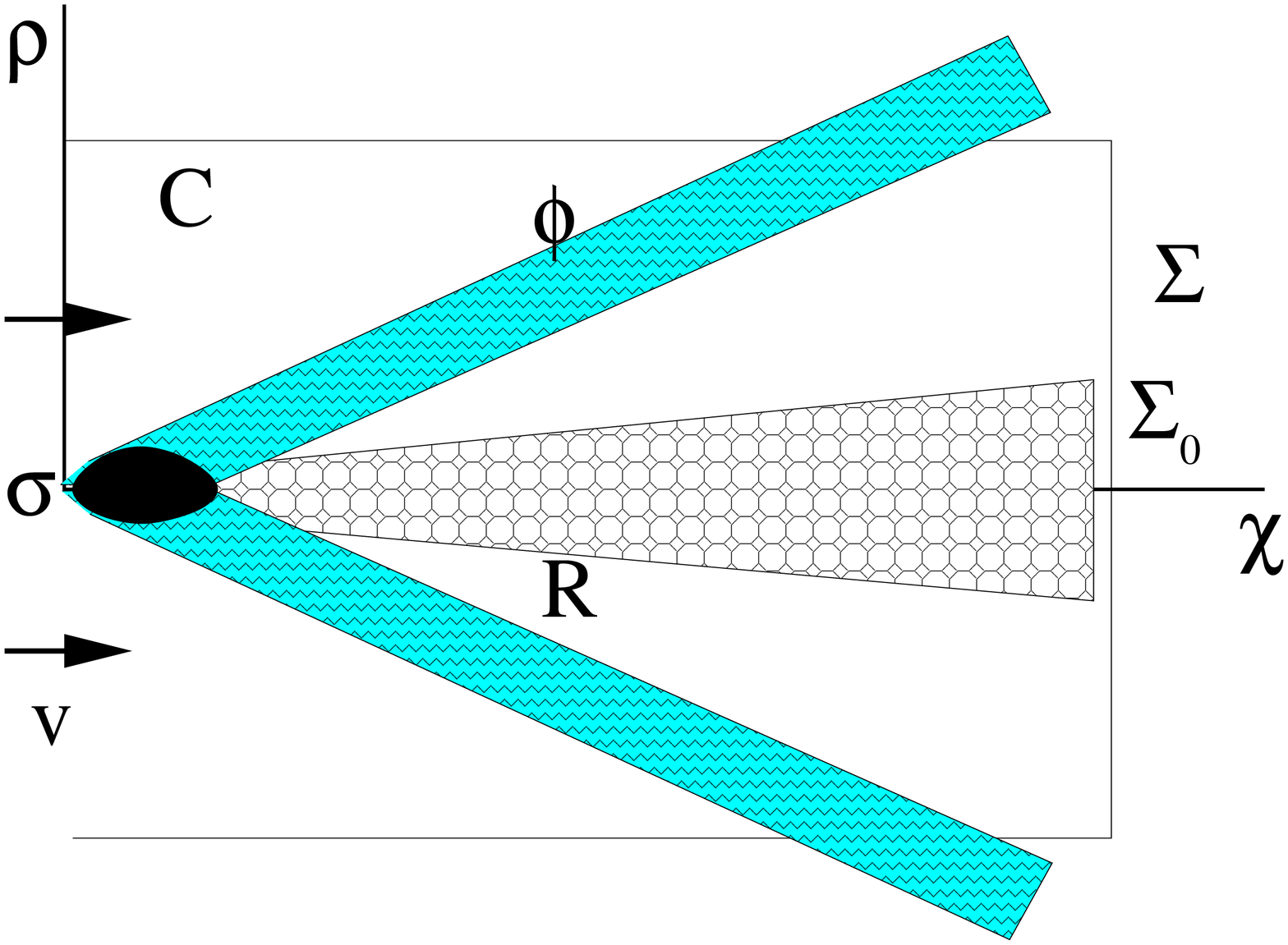}
 \caption{\label{fig_shocks}
Two schematic pictures of flow created by a jet going through
the fireball. 
}
 \end{figure}
  Fig.\ref{fig_shocks} explains a view of the process,
in a plane transverse to the beam.  Two oppositely
moving jets originate from   the hard collision point B.
 Due to strong quenching, the survival of the trigger
jet biases it to be produced close to the surface and to
 move outward. This  forces its companion to 
move inward through matter and to be maximally quenched.
The energy deposition starts at point B, thus a spherical sound wave
appears (the dashed circle in Fig.\ref{fig_shocks}a ). Further 
 energy deposition is along the jet line, and is propagating with a
 speed 
$v$ of the jet
till the leading parton is found at point A
at the moment of the snapshot.
The non-hydrodynamical core (solid region) serves as a source for
the hydrodynamic fields. 

The main prediction is that the shape of the jet passing through sQGP
drastically changes: most of
associated secondaries 
  fly preferentially to a very large angle with jet direction,
 $\approx 70$ degrees
 consistent with the Mach angle  $ cos\theta_M=v/c_s $
with a (time-averaged) speed of sound
 \be \bar c_s^{RHIC}={1/\tau}\int_o^\tau dt c_s(t)\approx .33 \ee
  
Antinori and myself~\cite{Antinori:2005tu} suggested to test it
  further
using  b-quark jets,
which can be tagged experimentally. As they get
less relativistic
  the Mach
angle should  shrink, till it  vanishes at the critical velocity
$v=c_s=1/\sqrt{3}$. Note that such
behavior of the cone is opposite
to what happens for gluon radiation, which predicts
 a wider distribution
 with $decreasing$
$v$.

Casalderrey and myself\cite{CS_variable} have shown, using 
conservation of adiabatic invariants, that fireball expansion 
should greatly enhance the sonic boom. The reason is similar to
enhancement
of a sea wave such tsunami as it goes onshore: see below.

Chaudhuri and Heinz \cite{Uli_conical} have numerically solved  (in
rapidity-independent 2+1 hydro) hydrodynamical equations
with the energy deposited by a jet. Although they see a shock
in coordinate space, they failed to see any peaks at the Mach angle
in spectra. See more on it in the next section and Heinz' talk.
On the other hand, 
 partonic cascades (with large cross sections appropriate for sQGP)
 have actually found quite clear signature of the conical flow, see
Ma  et al. \cite{MAMPT}.

 It turns out\footnote{It was pointed out to me by B.Jacak
recently.} that observations of the Mach cone is  routinely used
experimentally in  studies of strongly coupled
``dusty'' electromagnetic
plasmas : see e.g. \cite{Nosenko_dusty}, where one can see
double Mach cones, corresponding to both sound and ``shear''
modes.

\section{The conical flow: excitation of two modes}
\label{}
Let me start by explaining why the hydrodynamical approach is
able to predict only the $shape$ of the correlation function induced
by conical flow,  not its $amplitude$. 
 The reason is simple: in a region close to the 
jet the variation of energy density is so rapid that
hydrodynamics cannot be applied there. This  preclude us from 
predicting the amount of the {\em produced entropy} $dS/dx$ by the jet, 
as well as the fraction of energy  going into  sound and
``diffuson''
modes\footnote{It does not matter whether
hydro is or is nor linearized: thus Chaudhuri and Heinz\cite{Uli_conical}
could not possibly determine it as well. The only approach which does
not have this problem is AdS/CFT to be discussed below. }. 

  The jet energy deposited in a ``viscous volume'' 
 \be
\frac{E_{\rm lost}}{E_{\rm fluid}}\approx 
                       \frac{\frac{dE}{dx}\times\Gamma_s}{e\times\Gamma_s^3}
		       \approx 36-100 \gg 1 \, .
\ee
 is numerically large, and so a
jet is surrounded by its own small ``fireball''
(of size $\Gamma_s$ or more) where  variation 
of the thermodynamic quantities
 is too large for hydrodynamics to be applicable. 
Outside of this region, there is a domain where gradients are small enough
that viscous hydrodynamics can in principle be used, but
the behavior of the fluid is 
non-linear, dissipative, and possibly turbulent.
We will not discuss these complex regions and proceed to large
 distances where linearized hydro should work. 

 According
to the axial symmetry of the problem, the most general expression for 
the initial disturbance of the energy density and momentum in medium are
\be
\label{ini_cond}
\epsilon_{dt_0}(t=t_0,{\bf x})=e_0(x,r) \\ \nonumber
{\bf g}_{dt_0}(t=t_0,{\bf x})=g_0(x,r)\delta^{ix}
+{\bf \nabla} g_{1}(x,r) 
\ee 
The source functions $e_0(x,r)$ and $g_{1}(x,r)$ excite only
 the sound mode, while the remaining function
$g_0(x,r)$ excites the diffuson mode.
The particular value of these functions depends on the
interaction of the jet and the fluid in the near region. 

\begin{figure}[t]
\begin{center}
\includegraphics[width=6cm,height=10cm]{dNdydphipt_SL.epsi}
\includegraphics[width=7cm]{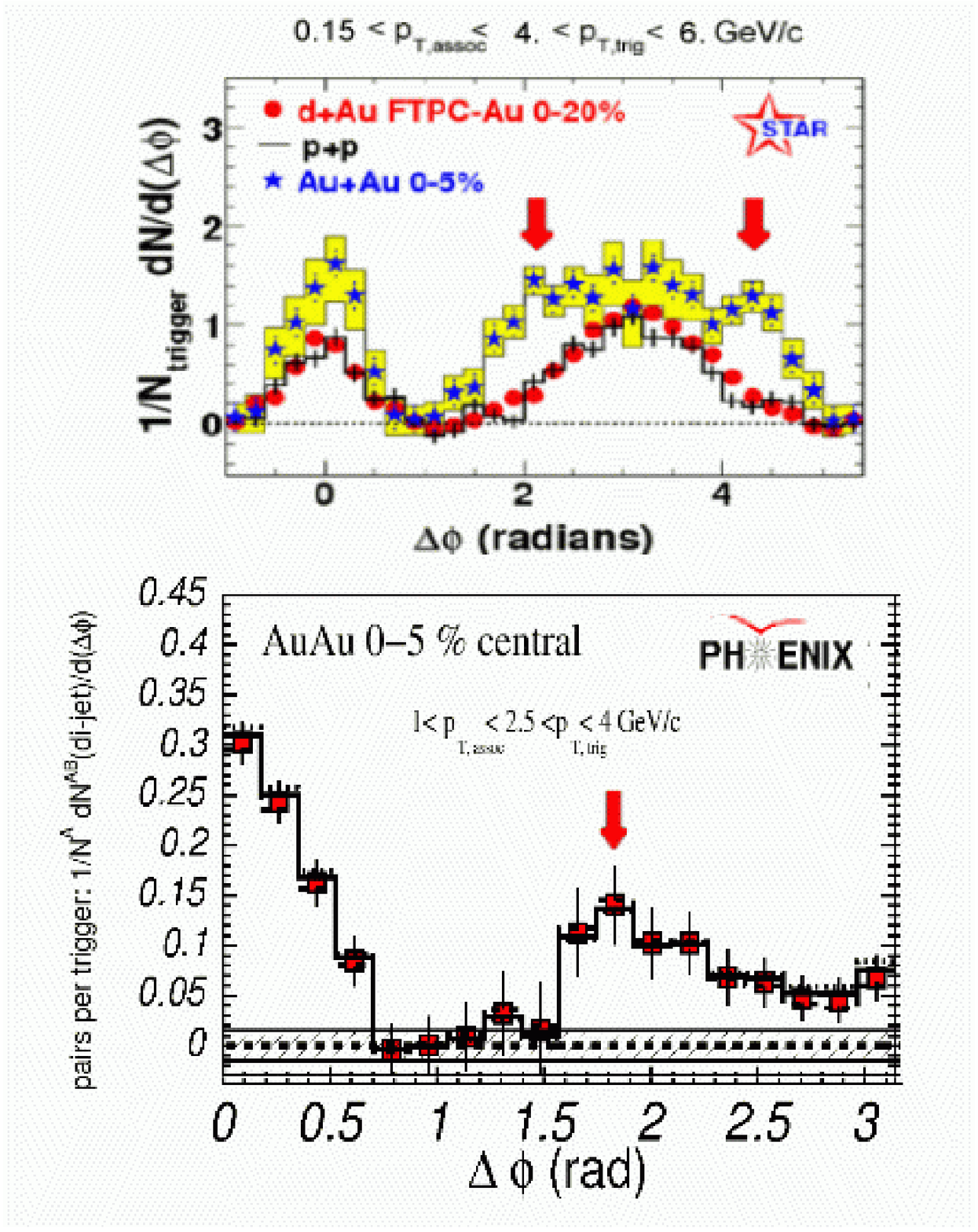}
\end{center}
\caption{
\label{corptdep}
Left: Associate yield dependence on associate $p_T$ 
for fixed source size $\sigma=0.75/T$
 , viscosity $\Gamma_s=0.1/T$, $t_j=8/T$, $t_f=10/T$,
and energy loss, $dE/dx=10 T^2$ (top) and $dE/dx=63 T^2$ 
(bottom). The label values for $dE/dx$ correspond to 
$T=200~\mbox{MeV}$.
 The three curves are for  
 $1 T<p_{t}<5 T$ (solid),
 $5 T<p_{t}<10 T$ (dotted),
($3\times$) $10 T<p_{t}<15 T$ (dashed),
($10\times$) $15 T<p_{t}<20 T$ (dashed-dotted).
(in the upper panel all the curves are rescaled further up
by a factor 10). No large angle correlation is observed
for $dE/dx=10 T^2$. For $dE/dx=63 T^2$
the position of the peak shifts toward $\pi$
for lower $p_T$.
Right: Experimental dihadron azimuthal distributions from
STAR (top) \cite{star_peaks} and 
PHENIX (bottom) \cite{phenix_peaks}
}
\end{figure}

The reader should consult the rather long paper \cite{CST}
for many details about calculated angular distributions
of secondaries, and its dependence on parameters involved.
Like for elliptic flow, the effect of conical flow grows
with $p_t$ of the secondaries and seem to get maximal at $p_t\sim
1-2\, GeV$. It is also strongly dependent on jet quenching
deposition $dE/dx$, as demonstrated in Fig.\ref{corptdep}(a).
But mostly it is sensitive to viscosity: since the conical
flow has larger gradients than elliptic one, one naturally may expect
that eventually this phenomenon will provide its best estimates.

\section{Expanding fireball and variable speed of sound}
 The
  speed
of sound is defined by $ c_s^2=dp/d\epsilon $
via the thermodynamical variables, and it is expected to change in wide
range during the process of heavy ion collisions. At early stages at
  RHIC
the matter is believed to be  in form of
quark-gluon plasma  (QGP), and thus
it is $c^2_{QGP}=1/3$. The next stage is the so called ``mixed
  phase''
in which the energy density is increasing much more rapidly than
  pressure,
so that $c^2$ decreases to rather small values at
 the so called ``softest point'', 
and then rise again to $c^2_{RG}=.2$ in the hadronic ``resonance
  gas''. 

Casalderrey and myself \cite{CS_variable}
  discuss issues related with it. The main result is
 that expansion and 
a decrease of the sound speed can significantly increase
the amplitude of the sound wave. 
The second issue is
what will happen if the QCD phase transition is of the 1-st order, so that
there is truly mixed phase and the  minimal value of $c^2$
is zero. 

 A motion of
any waves in weakly inhomogeneous medium  can be described
via geometric optics and
eikonal eqns for the $phase$ of the wave. Its $amplitude$ is much more
tricky to get, which can sometimes be addressed with the help of {\em adiabatic
  invariants} $I=\oint pdq$. Let me remind the reader
what conservation of $I$ means for  a basic example,
a harmonic oscillator with a slowly variable energy and frequency
($d log\omega(t)/d log t\ll 1$ etc). The typical momentum and amplitude
of oscillations scale as \be p\sim
\sqrt{E(t)},\,\, \, q\sim \sqrt{E(t)/\omega^2(t)}\ee
while
their product -- the adiabatic invariant $I\sim E(t)/ \omega(t)=const$ --
 remains constant.

		       Hydrodynamical equations
     for sound waves, in momentum representation in coordinates,
		       also form an oscillator,
and therefore we will use the adiabatic invariant to study the changes on the
   velocity field due to the expansion and variable speed of sound.
			   Unfortunately, a
	      simple substitution of a variable $c_s(t)$
into the equations of motion for perturbations of a static background
			   is inconsistent.
     One should instead find a correct non static solution of the
hydrodynamical equations and only then, using this solution as zeroth order,
      study first order perturbations such as sound propagation.
	 The numerical solutions for hydrodynamical equations
      have been done by a number of authors: but in all of them
      the flow and matter properties depend on several variables
		  and is too difficult to implement.

 Therefore, in order to study the effects of the variable speed of 
sound we have looked for the simplest example possible, in
which there is a nontrivial
time-dependent expansion
  but still  no spatial coordinates are involved, keeping the problem
homogeneous in space. The only way these goals can be 
achieved is by a Big-Bang 
gravitational
process, in which the space is created dynamically
by gravity. With such space available, the matter
can cool and expand at all spatial points in the same way.
For definiteness, we considered a liquid in  Robertson-Walker metric
of expanding Universe with time-dependent radius $R(t)$.
The sound wave eqns looks then as
\be
\label{second}
\partial^2_{\eta} \epsilon - c_s^2 \nabla \epsilon 
+ \epsilon\partial_{\eta} \left( (3c_s^2-1)\frac{R'}{R} \right ) + 
(3c_s^2-1)\frac{R'}{R} \partial_{\eta} \epsilon=0.
\ee
in proper time $\eta$.
Note that both corrections to the wave equations vanish for $c_s^2=1/3$ 
(the QGP value) and the rescaling factor 
 actually produces an amplitude growth, if the speed of sound reduces
below the ideal value 
$c^2_s <1/3$.
 The relevant
quantity for the final production of particles (the exponent
of the Cooper-Fry
integral) is  the velocity to temperature ratio $v^i/T$: according to
our calculations this ratio grows by about factor 3 by the time of
kinetic freezeout. 

Another important point of this paper is a study of what would happen
if the QCD phase transition would be the 1st order
and  the speed
of sound in the 
mixed phase
would vanish. We studied numerically and analytically what
 happens in this case, and found rather spectacular phenomenon: the
``frozen'' wave after the mixed phase splits into two halves, moving
in opposite directions. 

\begin{figure}[t] 
\begin{center}
\includegraphics[width=7cm]{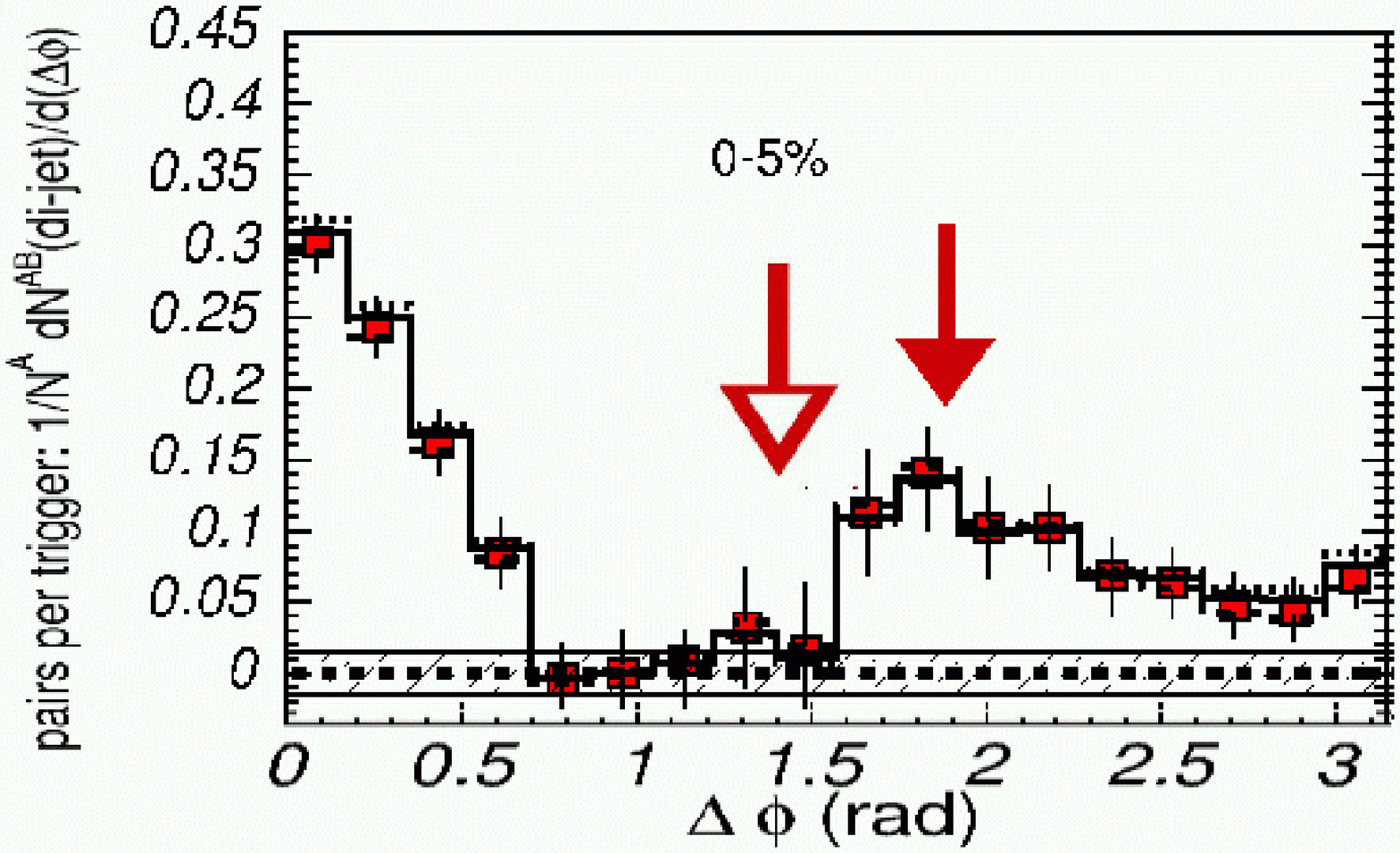}
\includegraphics[width=7cm]{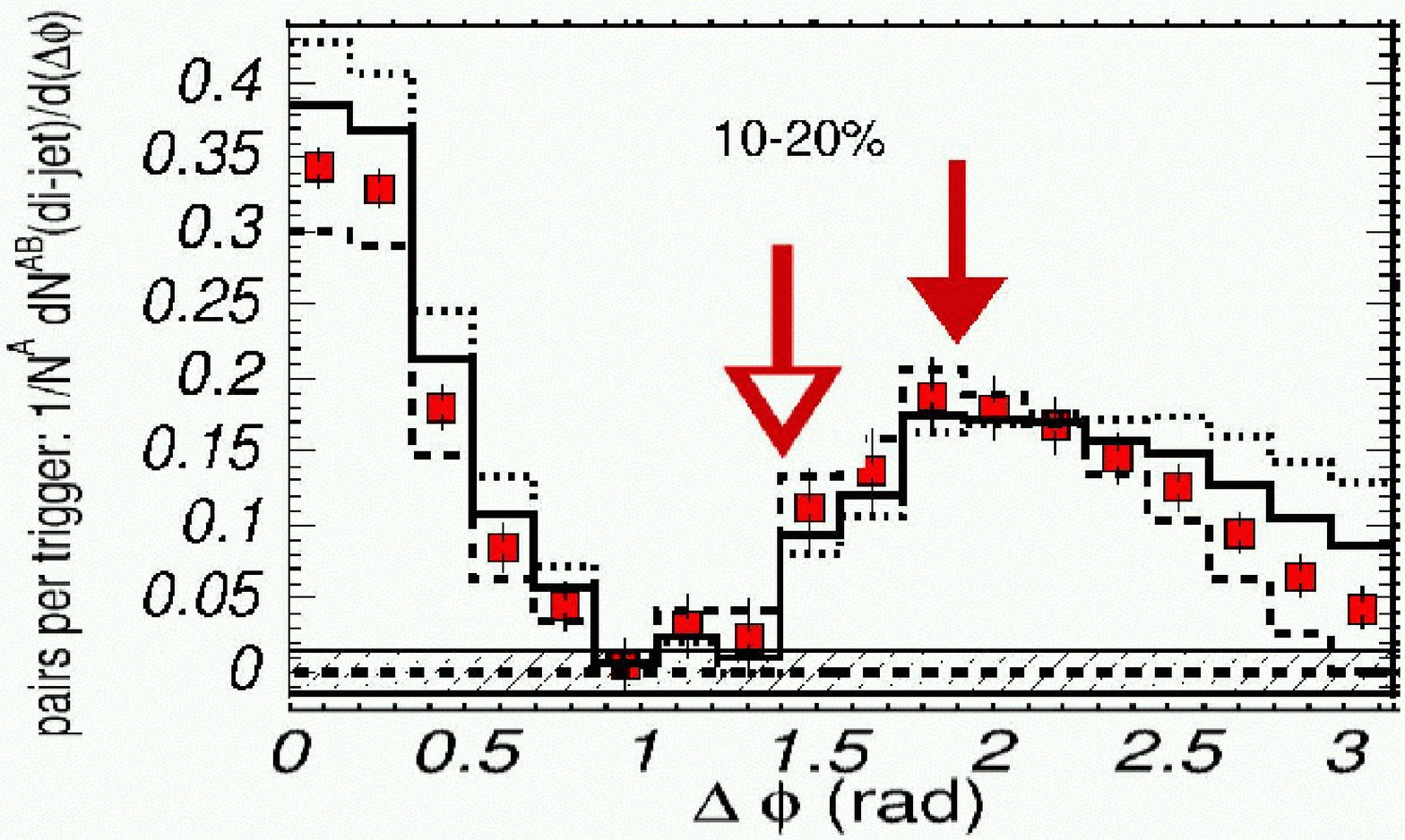}
\end{center}
\caption{
\label{exp_corr}
Azimuthal dihadron distributions normalized per trigger particle measured
by PHENIX \cite{phenix_peaks} for two different centralities 
($2.5 < p_T^{trigger} < 4 GeV$,$1 < p_T^{associated} < 2.5 GeV)$. 
The filled arrow indicates the position of the Mach cone. The empty
arrow our estimate for the position where
the cone from the reflected wave should appear. 
}
\end{figure} 

So, $if$ the QCD phase transition is
of the first order, the original wave gets split into direct and
reflected waves. At freezeout one would
 find the reflected wave on its way
 $to$ the origin, the opposite to the Mach direction of the
direct wave.  In terms of two particle 
azimuthal distribution it means the appearance of a {\bf second peak} 
at some angle
$\Delta \phi < \pi/2$. We return to it at the end of the next section.

\section{Experimental issues}
The first
experimental observations made at RHIC, first by the STAR collaboration
\cite{star_peaks} and then by PHENIX \cite{phenix_peaks},
are  2-particle correlation functions,
 for a trigger hard particle and ``companion''. Their relative
distributions in azimuthal angle shown above
(Fig.\ref{corptdep}b and Fig.\ref{exp_corr}) do indicate a
depletion of correlated particles in the direction of the quenched jet.
 A peak observed agrees with an angular position and shape in agreement
with hydro predictions.

The first experimental issue was  whether this effect is real,
as it only gets observed after elliptic flow subtraction.
It seems by now being resolved: see e.g. a 
talk by B.Cole \cite{Cole}, who
demonstrated 
that the shape of the correlation function is quite
independent of the direction of the jet in respect to the 
flow. Additionally one may use
a subset of the data, taken at the particular
 angle at which the contribution of the elliptic flow vanishes,
and still see  a clear minimum at
$\delta \phi=\pi$ and a large-angle peak.
 
 The next issue is whether one indeed see a conical picture
in geometrical sense, not just its projection onto the azimuthal
angle. The next observable after two-body correlation function is
naturally tree-particle correlations.
Although statistically it is very hard to do, PHENIX and STAR both
have emerging data on this, which were discussed at HP06 a lot. 
To my eye, those distributions look convincing, but perhaps
more work and further
scrutiny in collaborations is required to reach
the final conclusion on this.

 Let me at the end show
  Fig. \ref{exp_corr}  with samples of experimental correlation function as obtained
by PHENIX \cite{phenix_peaks}. 
The peak around 1.9 rad  (indicated by the 
filled arrow) corresponds to the Mach direction; this is the main
effect attributed to a conical flow. The  reflected wave should appear in 
the region $\Delta \phi< \pi/2$, and the empty arrow shows the estimated
place where the corresponding reflected peak should
 appear. Fig \ref{exp_corr} a) shows the most central collisions,
does not show any nonzero signal at that angle. Fig \ref{exp_corr} b), at
higher centrality, has a nonzero correlation function there, but  
it seems likely to be just a slope
of  a much broader peak.
 As  argued above, we see no indications for enhanced
correlations at the expected angle of the reflected wave,
and thus  the deconfining
phase transition {\em cannot be of the first order}.

 \begin{figure}
  \centerline{\includegraphics[width=5in,angle=270]{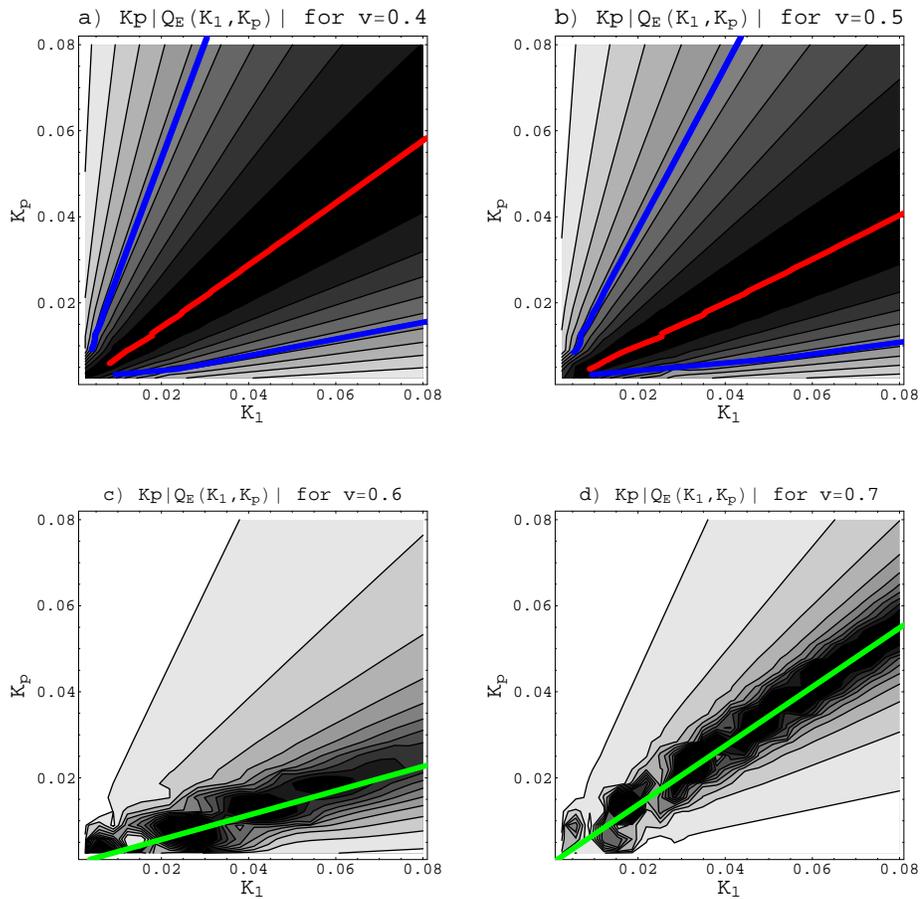}}
  \caption{(from \protect\cite{Friess:2006fk})
Contour plots of $K_\perp |Q^K_E|$ for various values of $v$ at low momenta.  The green line shows the Mach angle.  The red curve shows where $K_\perp |Q^K_E|$ is maximized for fixed $K = \sqrt{K_1^2+K_\perp^2}$.  The blue curves show where $K_\perp |Q^K_E|$ takes on half its maximum value for fixed $K$.}\label{fig:BoomFig}
 \end{figure}

\section{Conical flow is found via AdS/CFT}
  This section is about a development
which  was actually discussed in my talk
as a suggestion, but concluded after HP06.
 Being an optimist, I said then
that `` ...in the AdS/CFT framework one soon
be able to derive from graviton propagator
 what exactly is the flow pattern induced by
a heavy quark jet in a strongly coupled plasma''. 
But even for an optimist this prediction was confirmed
 surprisingly soon:  a $month$ later the
Princeton group \cite{Friess:2006fk}
 have completed such
(technically nontrivial) calculation. It involves a solution
of the (linearized) Einstein equation, 
which tells what is the metric correction $h_{\mu\nu}$ due to gravity of the
 string which is being dragged behind the jet (indicated by 
the graviton line in Fig.\ref{fig_wake}(b)).

Quite remarkably, when these authors analyzed harmonics
of the stress tensor at small momenta $k
\ll T$ (shown at the lower part In Fig.\ref{fig:BoomFig})
they have seen the narrow cone of emission along the Mach angle,
which is just the ``conical flow''! 
And, as one can see from two upper plots 
on this figure, for ``subsonic'' velocities $v=0.4,0.5<1/\sqrt{3}$,
that  pattern of emission is completely changed
and the Mach cone disappears, exactly as argued 
by Antoniri and myself \cite{Antinori:2005tu}.

As I already mentioned above, the duality between
gravity in the bulk and sound on the brane has been pointed out 
already by
Polikastro et al.So it was clear this would work: the
nontrivial output is the absolute normalization of the conical flow.
 And yet, a dynamical derivation  from first principles of
 a complex
hydro flow   is a  remarkable
achievement. 

Perhaps we are now ready to proceed from discussion of
sound (linearized eqns) to a full-blooded hydro, by going from
gravitons (linearized Einstein eqns) to a study of a full-blooded 
general relativity. 
  The ultimate goal would be to work out a {\bf complete
 gravity dual} to the
whole
RHIC collision process, in which
one should be able to derive thermalization and subsequent
hydrodynamical evolution from first principles, following 
 dynamically the gravity field of a  produced black hole~\cite{Nastase}. 
 Sin, Zahed and myself~\cite{Shuryak:2005ia}
further argued that exploding/cooling
fireball on the brane is dual to
 $departing$  black hole,
falling into the AdS center.
Janik and Peschanski~\cite{JP} indeed found that gravity dual to
  the Bjorken hydro
has a metric with a departing $horizon$. (However
they so far solved only vacuum Einstein eqns without any matter.)

{\bf Acknowledgments}  My collaborators, Derek Teaney
and especially Jorge Casalderrey,  worked out most of the
results I am presenting. The work is partially supported by the
grant of US DOE.


\end{document}